\documentclass{PoS}

\usepackage{wrapfig}
\usepackage{amssymb}

\title{The EPPS16 nuclear PDFs}

\ShortTitle{The EPPS16 nuclear PDFs}

\author{Kari J. Eskola \\
University of Jyvaskyla, Department of Physics, P.O. Box 35, FI-40014 University of Jyvaskyla, Finland \\
Helsinki Institute of Physics, P.O. Box 64, FI-00014 University of Helsinki, Finland \\
E-mail: \email{kari.eskola@jyu.fi}}

\author{Petja Paakkinen \\
University of Jyvaskyla, Department of Physics, P.O. Box 35, FI-40014 University of Jyvaskyla, Finland \\
Helsinki Institute of Physics, P.O. Box 64, FI-00014 University of Helsinki, Finland \\
E-mail: \email{petja.paakkinen@jyu.fi}}

\author{\speaker{Hannu Paukkunen} \\
University of Jyvaskyla, Department of Physics, P.O. Box 35, FI-40014 University of Jyvaskyla, Finland \\
Helsinki Institute of Physics, P.O. Box 64, FI-00014 University of Helsinki, Finland \\
Instituto Galego de F\'\i sica de Altas Enerx\'\i as (IGFAE), Universidade de Santiago de Compostela, E-15782 Galicia, Spain \\
E-mail: \email{hannu.paukkunen@jyu.fi}}

\author{Carlos A. Salgado \\
Instituto Galego de F\'\i sica de Altas Enerx\'\i as (IGFAE), Universidade de Santiago de Compostela, E-15782 Galicia, Spain \\
E-mail: \email{carlos.salgado@usc.es}}

\abstract{We report on EPPS16 - the first analysis of NLO nuclear PDFs where LHC p-Pb data (Z, W, dijets) have been directly used as a constraint. In comparison to our previous fit EPS09, also data from neutrino-nucleus deeply-inelastic scattering and pion-nucleus Drell-Yan process are now included. Much of the theory framework has also been updated from EPS09, including a consistent treatment of heavy quarks in deeply-inelastic scattering. However, the most notable change is that we no longer assume flavour-blind nuclear modifications for valence and sea quarks. This significantly reduces the theoretical bias. All the analysed data are well reproduced and the analysis thereby supports the validity of collinear factorization in high-energy collisions involving heavy nuclei. However, flavour by flavour, the uncertainties are still rather large.}

\FullConference{XXV International Workshop on Deep-Inelastic Scattering and Related Subjects\\
		3-7 April 2017\\
		University of Birmingham, UK}

\begin{document}

\section{Introduction}

\vspace{-0.3cm}
We have lately released a new set of nuclear PDFs, EPPS16 \cite{Eskola:2016oht}, which is meant to supersede our earlier parametrization EPS09 \cite{Eskola:2009uj}. The most important reforms with respect to EPS09, and also differences in comparison to the DSSZ \cite{deFlorian:2011fp} and nCTEQ15 \cite{Kovarik:2015cma} analyses, %which was the latest one before the EPPS16 release, 
can be seen from Table~\ref{tab:npdftable}. The new experimental input in EPPS16 consists of measurements of neutrino-nucleus deeply-inelastic scattering (DIS), Drell-Yan process in pion-nucleus collisions, and heavy-gauge-boson and dijet production in LHC p--Pb collisions. On the technical side, we have moved from the zero-mass variable flavour number scheme (ZM-VFNS) to the SACOT-$\chi$ general-mass variable flavour number scheme (GM-VFNS), and we have freed the flavour dependence of the quark nuclear modifications. Also, we no longer assign ad-hoc weights on data sets.

\vspace{-0.3cm}
{\small
\begin{table}[htb!]
\caption{Ingredients of the EPS09, DSSZ, nCTEQ15 and EPPS16 global nuclear-PDF analyses.}
\label{tab:npdftable}
\begin{center}
\vspace{-0.6cm}
\begin{tabular}{|c||c|c|c|c|}
\hline
& \textsc{eps09}  & \textsc{dssz} & \textsc{ncteq15} & \textsc{epps16} \\
\hline
{Order in $\alpha_s$}                   & LO \& NLO & NLO & NLO & NLO        \\
{Neutral current DIS $\ell$+A/$\ell$+d} & \checkmark  & \checkmark  & \checkmark     &  \checkmark     \\
{Drell-Yan dilepton p+A/p+d}            & \checkmark  & \checkmark  & \checkmark     &  \checkmark     \\
{RHIC pions d+Au/p+p}                   & \checkmark  & \checkmark  &  \checkmark  &  \checkmark                 \\
{Neutrino-nucleus DIS}                  &             & \checkmark  &         & \checkmark            \\
{Drell-Yan dilepton $\pi$+$A$}          &             &             & &  \checkmark                     \\                                            
{LHC p+Pb jet data   }                  &             &             & &  \checkmark              \\
{LHC p+Pb W, Z data  }                  &             &             & &  \checkmark               \\
& &                           &  & \\
{$Q$ cut in DIS}                        & $1.3 \, {\rm GeV}$ & $1 \, {\rm GeV}$       & $2 \, {\rm GeV}$   & $1.3 \, {\rm GeV}$  \\
{datapoints}                            & 929     & 1579                  & 708      & 1811             \\
{free parameters}                       & 15      & 25                  & 16       & 20             \\
{error analysis}                        & Hessian & Hessian                  & Hessian  & Hessian        \\
{error tolerance $\Delta \chi^2$}       & 50      & 30                  & 35        & 52           \\
{Free proton baseline PDFs}             & {\textsc{cteq6.1}} & {\textsc{mstw2008}}       & {\textsc{cteq6m}-like} & {\textsc{ct14}} \\
{Heavy-quark treatment}                 & ZM-VFNS   & GM-VFNS                & GM-VFNS & GM-VFNS            \\
{Flavour separation}                     &           &                & some & \checkmark \\
{Weight data in $\chi^2$}               &      yes  & no                & no & no \\
{Reference}                             & \textcolor{blue}{{\small {JHEP 0904 065}}}  & \textcolor{blue}{{\small {PR D85 074028}}} & \textcolor{blue}{\small{{PR D93 085037}}} & \textcolor{blue}{\small{{EPJ C77 163}}}  \\
\hline
\end{tabular}
\end{center}
\end{table}
}

\begin{wrapfigure}{r}{0.43\textwidth}
\vspace{-1.6cm}
\includegraphics[width=0.43\textwidth]{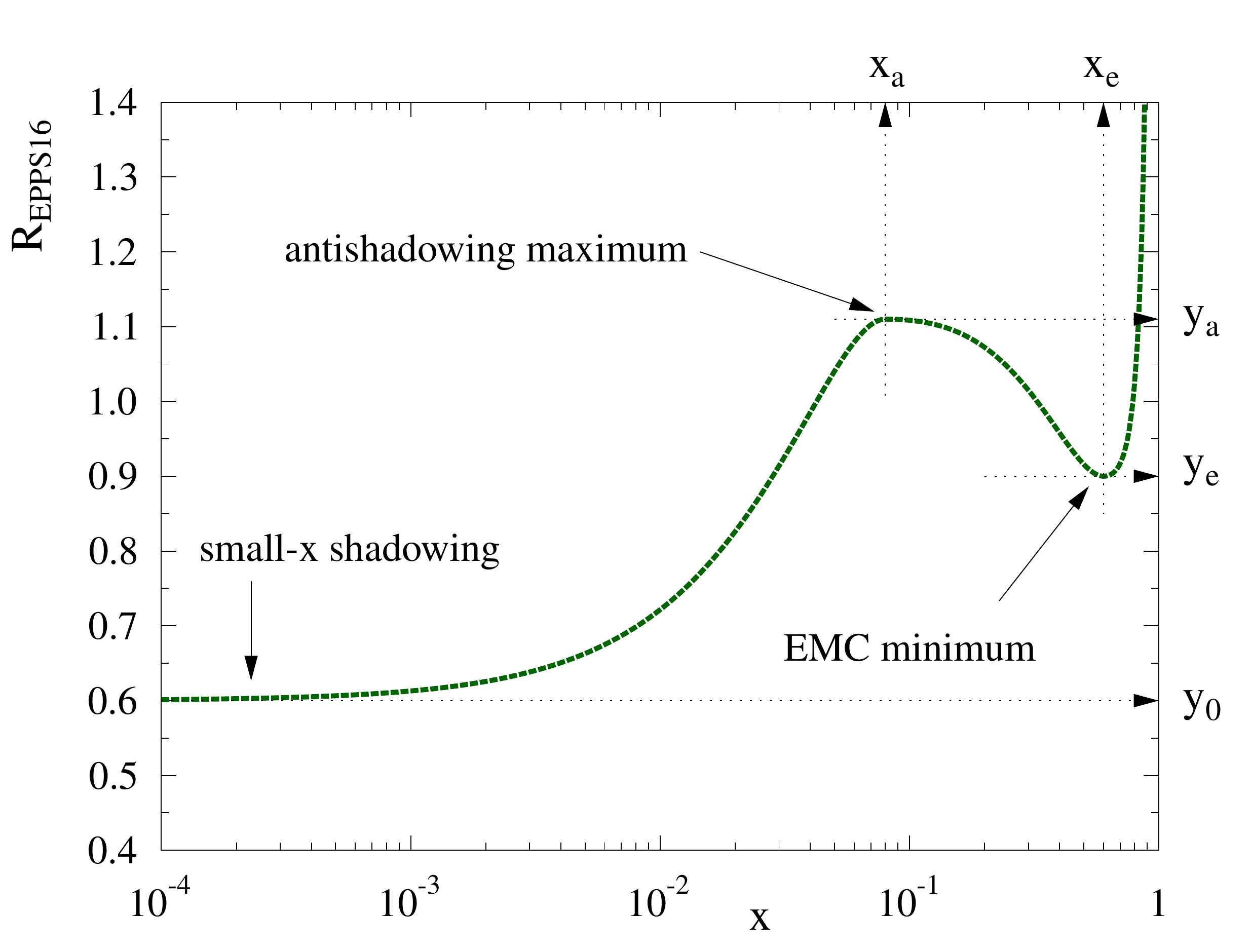} \\
\includegraphics[width=0.43\textwidth]{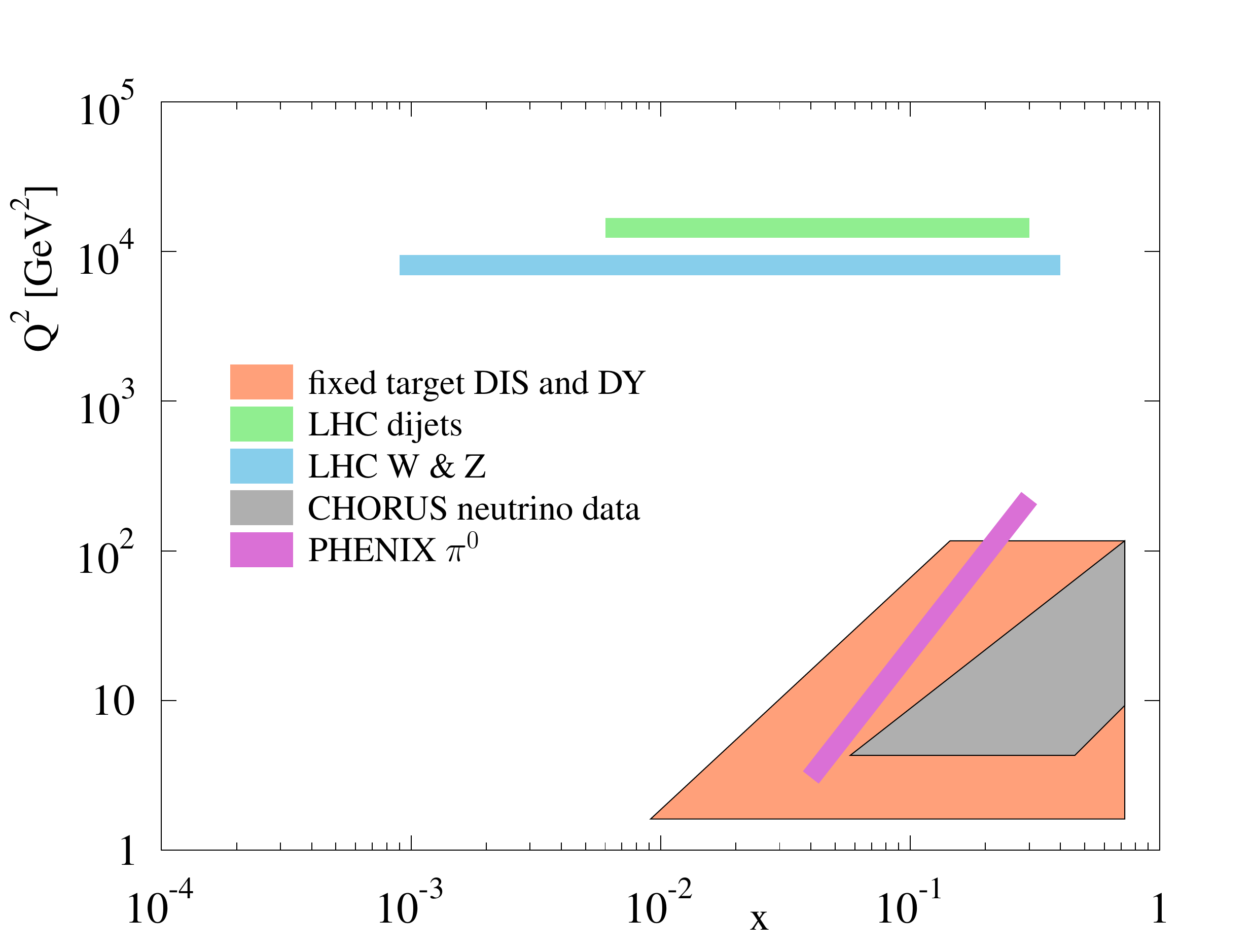}
\caption[]{\textit{Upper panel}: A typical form of the fit function in the EPPS16 analysis. \textit{Lower panel}: The input data used in the EPPS16 analysis in the $x,Q^2$ plane. Figures from Ref.~\cite{Eskola:2016oht}.}
\label{fig:schema}
\end{wrapfigure}

\vspace{-0.4cm}
As in EPS09, we define the bound-proton PDFs $f_i^{{\rm p}/A}(x,Q^2)$ for flavour $i$ as
\vspace{-0.1cm}
\begin{equation}
f_i^{{\rm p}/A}(x,Q^2) \equiv R_i^A(x,Q^2) f_i^{{\rm p}}(x,Q^2),
\end{equation}
where $f_i^{\rm p}(x,Q^2)$ is the free-proton PDF (here, CT14NLO \cite{Dulat:2015mca}), and $R_i^A(x,Q^2)$ the nuclear modification which we parametrize at the charm-mass threshold $Q^2=m_c^2$. The functional form we have in mind is similar to that shown in the upper panel of Figure~\ref{fig:schema}: at small $x$ we would expect shadowing ($R_i^A<1$), followed by antishadowing at mid $x$ ($R_i^A\gtrsim 1$), and then an EMC effect at large $x$ ($R_i^A<1$). We parametrize the nuclear mass-number ($A$) dependence at $x\rightarrow 0$, $x=x_a$, $x=x_e$ (see Figure~\ref{fig:schema}, upper panel) at the parametrization scale as
\vspace{-0.1cm}
\begin{equation}
R_i^A(x) = R_i^{A=12}(x) \left(\frac{A}{12} \right)^{\gamma_i(x) \left[R_i^{A=12}(x) - 1\right]},  \gamma_i(x)>0.
\end{equation}
\vspace{-0.1cm}
This, by construction, forces larger nuclear effects for larger nuclei. Unlike in EPS09 where we set $ R_{u_{\rm V}}(x,Q^2_0) = R_{d_{\rm V}}(x,Q^2_0)$ and $R_{\overline{u}}(x,Q^2_0) = R_{\overline{d}}(x,Q^2_0) = R_{\overline{s}}(x,Q^2_0)$, we now allow all the valence and light sea quarks to have their own independent nuclear modifications. 

\vspace{-0.5cm}
\section{Experimental data and their treatment}

\vspace{-0.3cm}
In comparison to EPS09, the data input in our new analysis has undergone a significant increase in both amount and variety. As illustrated in Figure~\ref{fig:schema} (lower panel), the new LHC p--Pb data probe the nuclear PDFs in a completely different $(x,Q^2)$ region than the older data. We have also changed the way how we use the old fixed-target DIS measurements: In EPS09 we included the data that the experiments had corrected for the unequal amount of protons and neutrons in the target. In EPPS16 we use the original, uncorrected data to avoid the potential bias caused by the external corrections and to gain more sensitivity to the flavour separation. The neutrino data in our fit come from the CHORUS collaboration \cite{Onengut:2005kv} with the full information on bin-to-bin correlated systematics. Following the ideas presented in Ref.~\cite{Paukkunen:2013grz}, the neutrino cross-section data are always normalized to the integrated cross sections at given neutrino-beam energy. This helps in reducing the theoretical as well as experimental uncertainties while sill retaining a clear sensitivity to the nuclear modifications. The LHC p--Pb data are always incorporated as forward-to-backward ratios. Again, this is to make the constraints more robust against theoretical and experimental uncertainties. During the fit, the LHC observables are evaluated with the aid of precomputed look-up tables. This enables a fast and accurate NLO-level treatment ``on the fly''.

\vspace{-0.5cm}
\section{Results}

\begin{figure}
\centering
\vspace{-0.8cm}
\includegraphics[width=0.95\linewidth]{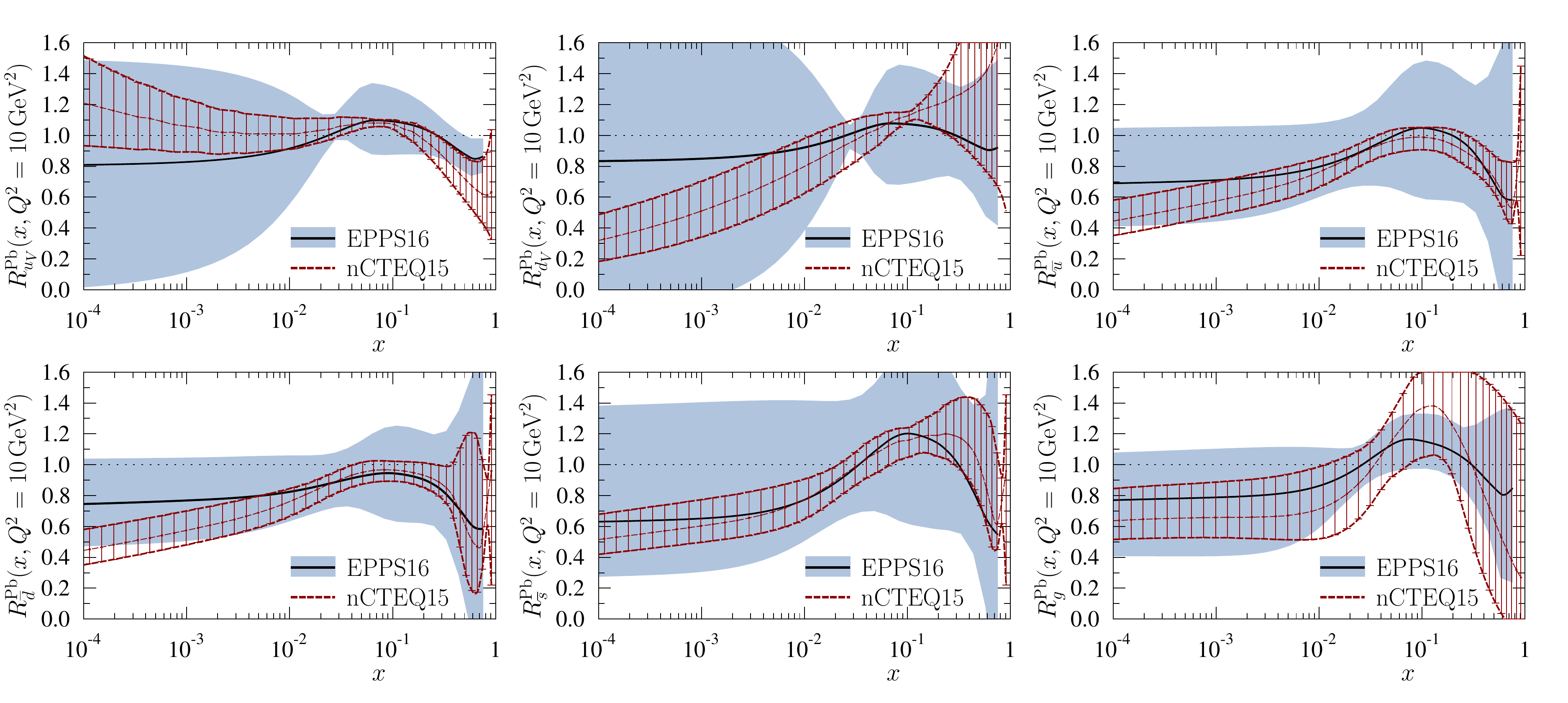}
\vspace{-0.3cm}\caption{The final EPPS16 nuclear modifications (black curves with blue error bands) at $Q^2=10 \, {\rm GeV}^2$ compared to the nCTEQ15 results (red bars). Figure taken from Ref.~\cite{Eskola:2016oht}.}
\label{fig:EPPS16andCTEQ}
\end{figure} 

\begin{wrapfigure}{r}{0.43\textwidth}
\vspace{-0.9cm}
\includegraphics[width=0.43\textwidth]{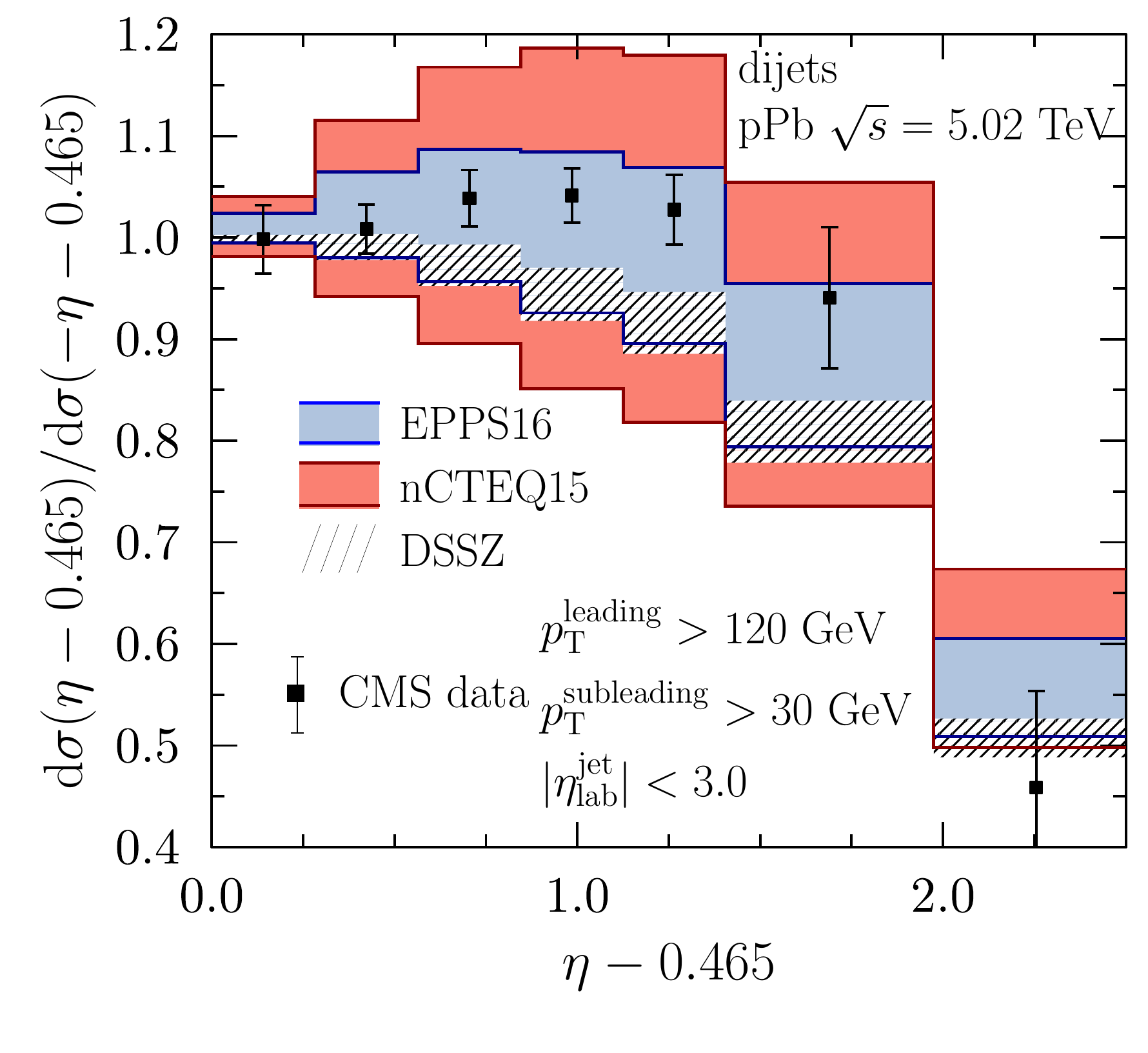}
\vspace{-1.0cm}\caption[]{The CMS dijet data \cite{Chatrchyan:2014hqa} compared with EPPS16, nCTEQ15 and DSSZ nuclear PDFs. Figure from Ref.~\cite{Eskola:2016oht}.}
\label{fig:dijet}
\end{wrapfigure}

\vspace{-0.3cm}
The final EPPS16 nuclear modifications at $Q^2=10 \, {\rm GeV}^2$ are presented in Figure~\ref{fig:EPPS16andCTEQ} where we also compare them to the corresponding nCTEQ15 results. The error bands shown are 90\% confidence levels, based on hypothesis testing. In the case of EPPS16 they correspond to a global tolerance of $\Delta \chi^2 = 52$. We have explicitly checked that the results are very similar to those with a dynamic tolerance criterion \cite{Martin:2009iq}. The central values of up and down valence-quark modifications in EPPS16 are mutually very alike. This in contrast to the nCTEQ15 results where the two show a quite different behaviour. We believe these differences originate from the facts that the nCTEQ fit did use the isoscalarized DIS data and that it did not include neutrino DIS data. The form of the fit functions may play a role as well. Also the sea-quark modifications come out quite similar in EPPS16 for all the flavours, even though they were free at small and mid $x$. The strange-quark modification is clearly less constrained, though. Overall, the nCTEQ15 uncertainties appear much smaller for the sea quarks. The basic reason for this is that nCTEQ15 did not allow any flavour freedom for the sea quarks. For the gluons, the nCTEQ15 uncertainties are clearly larger than those of EPPS16. Here, the role of the new CMS p--Pb dijet measurements \cite{Chatrchyan:2014hqa}, compared in Figure~\ref{fig:dijet} with the EPPS16, nCTEQ15 and DSSZ \cite{deFlorian:2011fp} nuclear PDFs, is essential: While the nCTEQ15 prediction encloses these data very well, the uncertainties are much larger than those of the EPPS16 analysis, in which these data are now included. %The effect of these measurements on nCTEQ15 would presumably be to bring the gluon uncertainty to a similar level with EPPS16.
The DSSZ PDFs clearly undershoot the dijet data for its very mild gluon nuclear effects.

\begin{figure}
\center
\vspace{-0.8cm}
\includegraphics[width=0.45\textwidth]{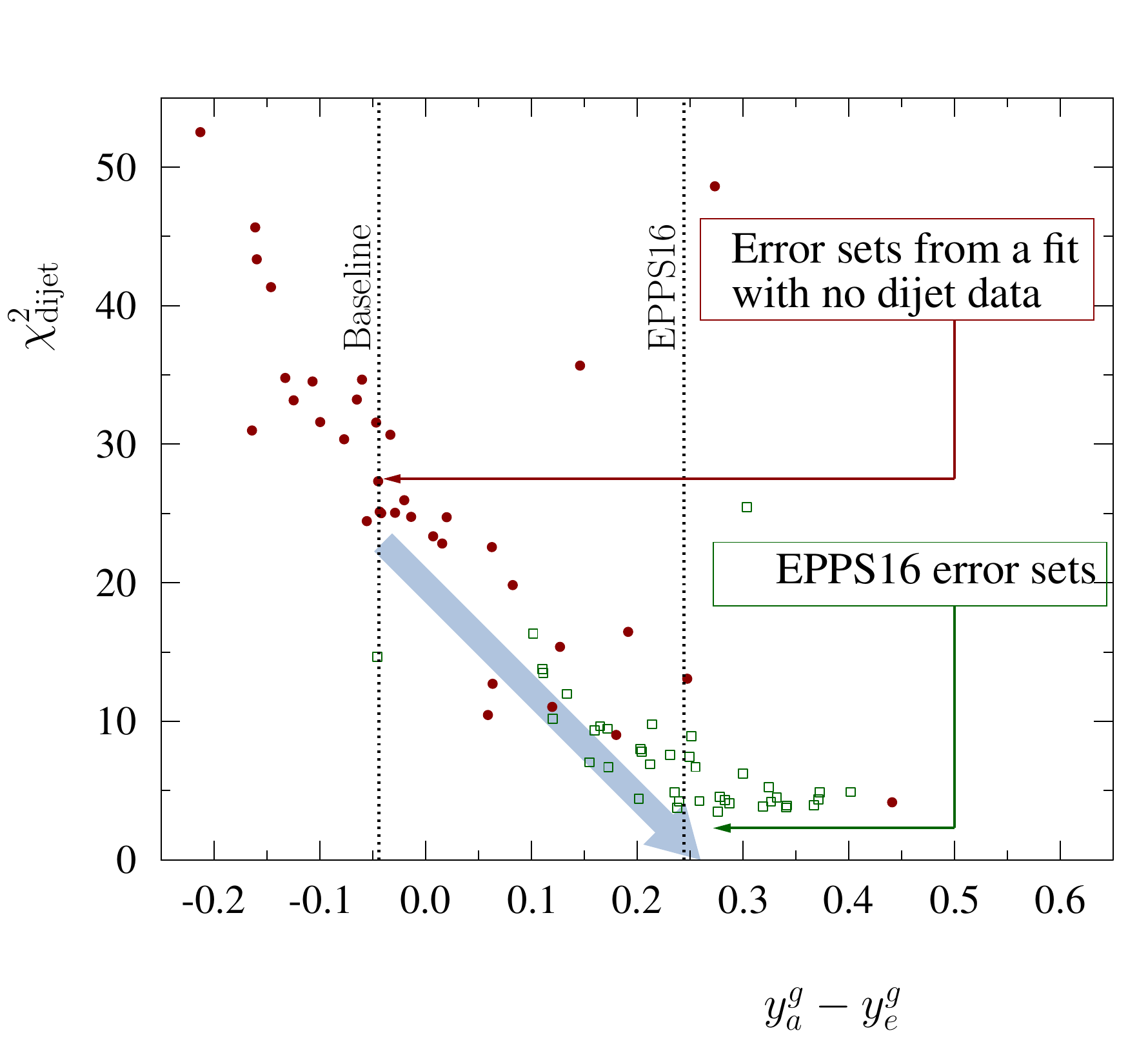}
\includegraphics[width=0.45\textwidth]{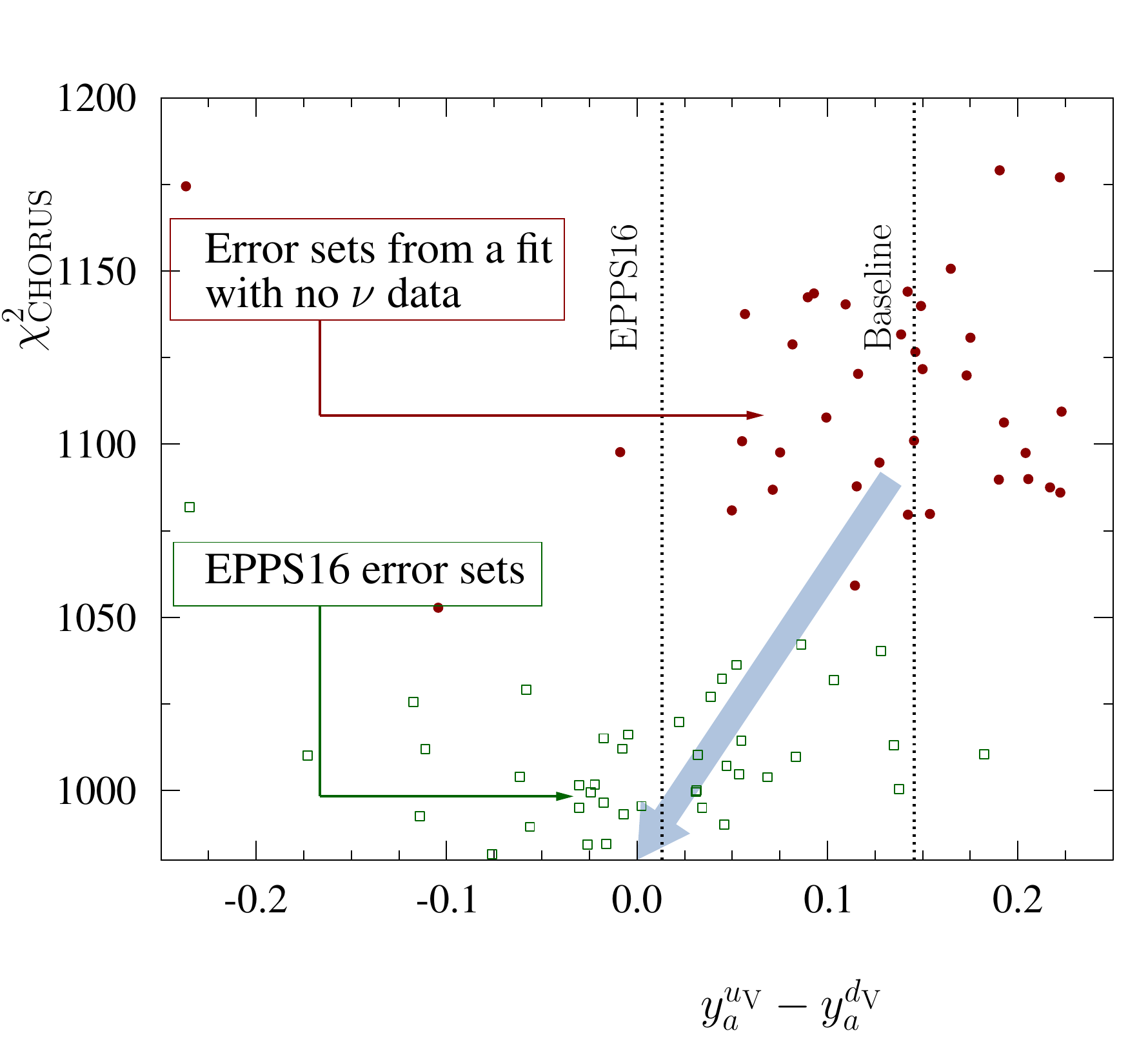}
\vspace{-0.3cm}\caption[]{\textit{Left:} The $\chi^2$ contribution of the CMS dijet data as a function of gluon fit-parameter combination $y_a^{\rm g}-y_e^{\rm g}$ (see Figure~\ref{fig:schema}) from a fit excluding (red) and including (green) these data. \textit{Right:} As the left panel but for the CHORUS neutrino data \cite{Onengut:2005kv} as a function of the valence-quark parameter combination $y_a^{{\rm u}_V}-y_a^{{\rm d}_V}$ (see Figure~\ref{fig:schema}). Figure from Ref.~\cite{Eskola:2016oht}.}
\label{fig:effects}
\end{figure}

The importance of the dijet data within the EPPS16 analysis is illustrated in Figure~\ref{fig:effects} (left panel) which shows the $\chi^2$ contribution of these data as a function of the ``strength'' of the gluon EMC effect (the difference between the antishadowing maximum and EMC minimum, see Figure~\ref{fig:schema}). If these data are excluded from the fit, the central result (called Baseline in Figure~\ref{fig:effects}) indicates no EMC effect for the gluons, but once these data are included, the $\chi^2$ decreases significantly and a clear EMC effect develops. Similarly significant is the role of CHORUS neutrino DIS data \cite{Onengut:2005kv}. If these data are not included, the central result indicates a clear difference between $R^A_{u_{\rm V}}$ and $R^A_{d_{\rm V}}$, as shown in Figure~\ref{fig:effects} (right panel). The effect of these data is to shift the central result to a region where $R^A_{u_{\rm V}} \approx R^A_{d_{\rm V}}$ with around a hundred-unit decrease in $\chi^2$. 

\vspace{-0.5cm}
\section{Summary}

\vspace{-0.4cm}
We have described the EPPS16 analysis of nuclear PDFs. This is the first analysis that directly includes constraints from the LHC p--Pb data. The most important LHC data set is currently the CMS dijet sample that helps in constraining the large-$x$ gluons. The neutrino DIS data are included as ratios to the integrated cross sections and these data provide important constraints for the valence quarks. The EPPS16 parametrization is also the first one to consider a full flavour decomposition for the quarks thereby significantly reducing the theoretical bias. The analysis supports the validity of collinear factorization in nuclear collisions in a largest $(x,Q^2)$ domain examined to date. However, the uncertainties are still large flavour by flavour.

\vspace{-0.5cm}
\section*{Acknowledgments}

\vspace{-0.4cm}
We have received funding from the Academy of Finland, Project 297058 of K.J.E; the European Research Council grant HotLHC ERC-2011-StG-279579 ; Ministerio de Ciencia e Innovaci\'on of Spain and FEDER, project FPA2014-58293-C2-1-P; Xunta de Galicia (Conselleria de Educacion) - H.P. and C.A.S are part of the Strategic Unit AGRUP2015/11. The financial support from the Magnus Ehrnrooth Foundation is acknowledged by P.P.

\end{document}